# Pure second harmonic current-phase relation in spin filter Josephson junctions

Avradeep Pal*, Z. H. Barber, J. W. A. Robinson, and M. G. Blamire

Department of Materials Science and Metallurgy, University of Cambridge, 27 Charles Babbage Road, Cambridge, CB3 0FS, United Kingdom

*e-mail: ap638@cam.ac.uk

**Abstract:** Higher harmonics in current phase relations of Josephson Junctions (JJs) are predicted to be observed when the first harmonic is suppressed. Conventional theoretical models predict higher harmonics to be extremely sensitive to changes in barrier thickness, temperature, etc. In contrast, experiments with JJs incorporating a spin dependent tunnelling barrier reported here reveal a current phase relation for highly spin polarized barriers which is purely 2$^{nd}$ harmonic in nature, and this is insensitive to changes in barrier thickness. This observation is consistent with recent theoretical predictions of a robust 2$^{nd}$ harmonic current phase relation for certain JJs with ferromagnetic barriers and implies that the standard theory of Cooper pair transport through tunnelling barriers is not applicable for spin dependent tunnelling barriers.

## Introduction

The supercurrent through a Josephson Junction (JJ) is conventionally described by means of a current-phase relation (CPR) $I = I_c \sin\varphi$ where $I_c$ is the critical current and $\phi$ is the junction phase-difference. More generally, the CPR can be expressed as $I = \sum_{n\geq 1} I_{c_n} \sin(n\varphi)$ [1,2] and, under circumstances in which the first harmonic is suppressed (for example at a 0-π transition), the 2$^{nd}$ harmonic may become detectable[3,4]. Recently it has been predicted that, in JJs with ferromagnetic barriers and asymmetric spin-active interfaces, the CPR should be inherently dominated by the 2$^{nd}$ harmonic as a consequence of the coherent transport of two triplet pairs[5,6]. Here we report measurements of tunnel junctions with ferromagnetic GdN barriers in which the period of the magnetic field modulation of $I_c(H)$ halves at the onset with thickness corresponding to a large spin polarisation of the tunnelling. For the largest thicknesses $I_c(H)$ tends to a perfect half-period Fraunhofer pattern corresponding to a pure 2nd harmonic CPR[7].

A dominant 2nd harmonic in the CPR can be manifested as half-integer Shapiro steps and magnetic interference patterns $I_c(H)$ with half the expected period[7]. While there have been experimental reports of half-integer Shapiro steps near to 0 to $\pi$ transitions in S/ F/ S JJs with both weak[3] and strong[4] diffusive F layers, evidence of a 2nd harmonic in CPR has not been reported in $I_c(H)$ patterns, although the periodicity of $I_c(H)$ is well known to be the most unambiguous probe of CPR.

Gadolinium Nitride (GdN) is one of the few known ferromagnetic insulators and has previously been shown by our group to yield high quality superconducting tunnel junctions when placed between NbN electrodes[8]; normal-state measurements have demonstrated spin filter behaviour with high spin polarisation $P$[9]. The devices reported here show a spin polarisation exceeding 80% which is higher than previously reported; for fabrication and measurement details see the methods section.

## Results

The $I_c(H)$ pattern of a JJ with a magnetic barrier is distorted by the field-dependent flux arising from the barrier magnetism[10] and so analysis requires the magnetic state of the barrier to be understood. In Fig. 1 we show typical $I_c(H)$ patterns of a GdN JJ with a high $P$ (approx. 89% at 4.2K, procedure of calculation of P is provided in supplementary information section) where the sequence of applied fields is split into three stages for clarity. Fig. 1(a) shows the initial field application sequence, which corresponds to the virgin curve of the corresponding GdN magnetisation hysteresis loop $M(H)$, in which the initially unmagnetised GdN layer is saturated by subjecting it to higher magnetic fields. Fig. 1(b) shows the behaviour of the critical current when the field is reduced from positive to negative saturation fields and Fig. 1(c) shows the return branch to positive saturation.

Several features are observed in Fig.1 which are distinctly different from $I_c(H)$ patterns of conventional JJs. Firstly, the role of field history is evident as the maximum critical current of the central, or first, lobe is hysteretically shifted from zero field in (b) and (c), indicating that the flux arising from the magnetic moment of the GdN barrier needs to be offset by an externally applied field in order to obtain the zero-flux maximum $I_c$[10]. Secondly, the ratio $I_{max2}/I_{max1}$ is significantly lower than the value of 0.21 expected for a conventional Fraunhofer $I = I_C|sinx/x|$ dependence. This indicates the relative suppression of $I_c$ in the second lobe; a feature that was noted earlier[8] and was attributed to enhancement of the pair current in domain walls owing to large area devices. Finally, the field widths of the lobes of the $I_c(H)$ pattern are not constant so that $H_{2+} \neq H_{2-} \neq H_1$. These distortions from the behaviour expected for a perfect Fraunhofer pattern are due to the fact that thin film GdN has a low coercivity of 20-50 Oe[9] and, because of their large area, our devices have multiple magnetic domains, and micro-magnetic structure plays an important role in multi-domain devices in distorting the $I_c(H)$ patterns[10] especially in the switching regions. Because GdN has a high remnant moment, the second lobe in Fig. 1(b) marked as $H_{2+}$, or its equivalent on the negative field side in Fig. 1 (c) should be least distorted by the changes in flux arising from the barrier micro-magnetics.

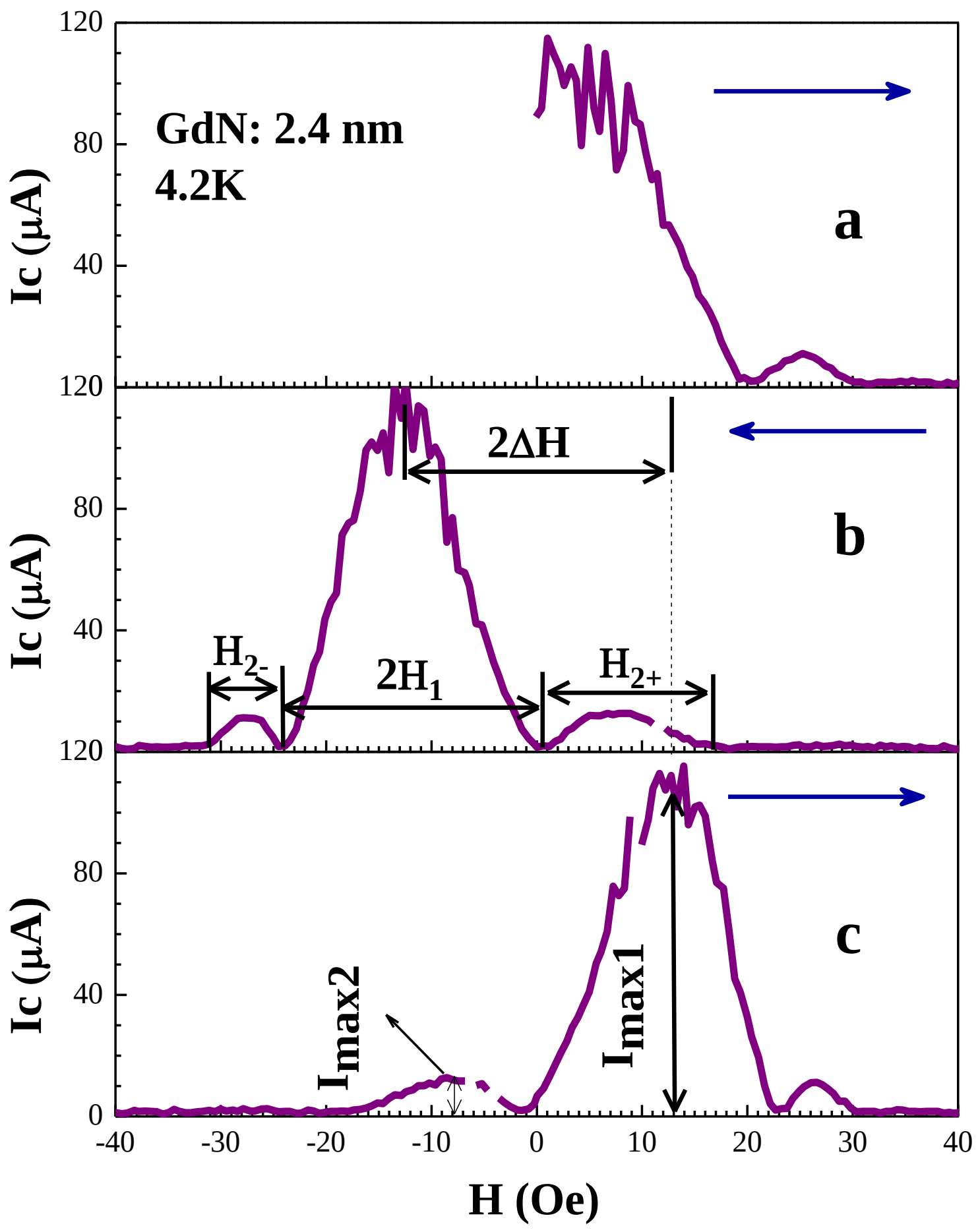

**FIG 1 –** Behaviour of critical current with externally applied in plane magnetic field broken into 3 parts showing the direction of field sweeps in each case. $\Delta \boldsymbol{H}$ is the hysteresis, $\boldsymbol{H_1}$is the magnitude of first lobe, $\boldsymbol{H_{2+}}$ is the second lobe magnitude obtained after saturation of GdN, $\boldsymbol{H_{2-}}$ is second lobe magnitude with the effect of micro-magnetics, $\boldsymbol{I_{max1}}$ and $\boldsymbol{I_{max2}}$ are the maximum critical currents in the first lobe and second lobes respectively.

In order to assess the extent of any such distortion, we model the field dependence of the magnetic moment of a hysteresis loop of GdN by a function $M(H) = \frac{a}{(1+be^{-k(H-C)})} - \frac{a}{2}$, where $\frac{a}{2}$ is the saturation moment of the GdN at a particular temperature and $b, c, k$ are fitting parameters. A fit to a measured $MH$ loop using this function is provided in the supplementary information section. For magnetic barriers, since the barrier magnetic moment couples with the externally applied magnetic field, the relation for critical current variation with magnetic field can be expressed as:

$$I = I_C \frac{Sin\{\frac{\pi(\varphi + \varphi_B)}{\varphi_0}\}}{\{\frac{\pi(\varphi + \varphi_B)}{\varphi_0}\}} = I_C \frac{Sin[\pi\left\{\left(\frac{x}{F}\right) + (\frac{\varphi_B}{\varphi_0})\right\}]}{\pi\{\left(\frac{x}{F}\right) + (\frac{\varphi_B}{\varphi_0})\}}$$

$$\frac{\varphi_B}{\varphi_0} = (d \times L) \times \left(\frac{4\pi}{d \times L^2}\right) \times \left(\frac{a}{(1 + be^{-k(H-C)})} - \frac{a}{2}\right) \times \frac{1}{2.06783461 \times 10^{-7}}$$

Where the flux quantum $\varphi_0 = 2.06783461 \times 10^{-7}$ G·cm², $\varphi_B$ is the flux due to barrier moment, $\varphi$ is flux due to externally applied magnetic field, $H$ is the externally applied magnetic field in Gauss, $F$ is the magnitude of magnetic field corresponding to one flux quantum for a particular device geometry, $d$ is the thickness of GdN layer and $L$ is the length of device edge perpendicular to magnetic field. All calculations are done in CGS units.

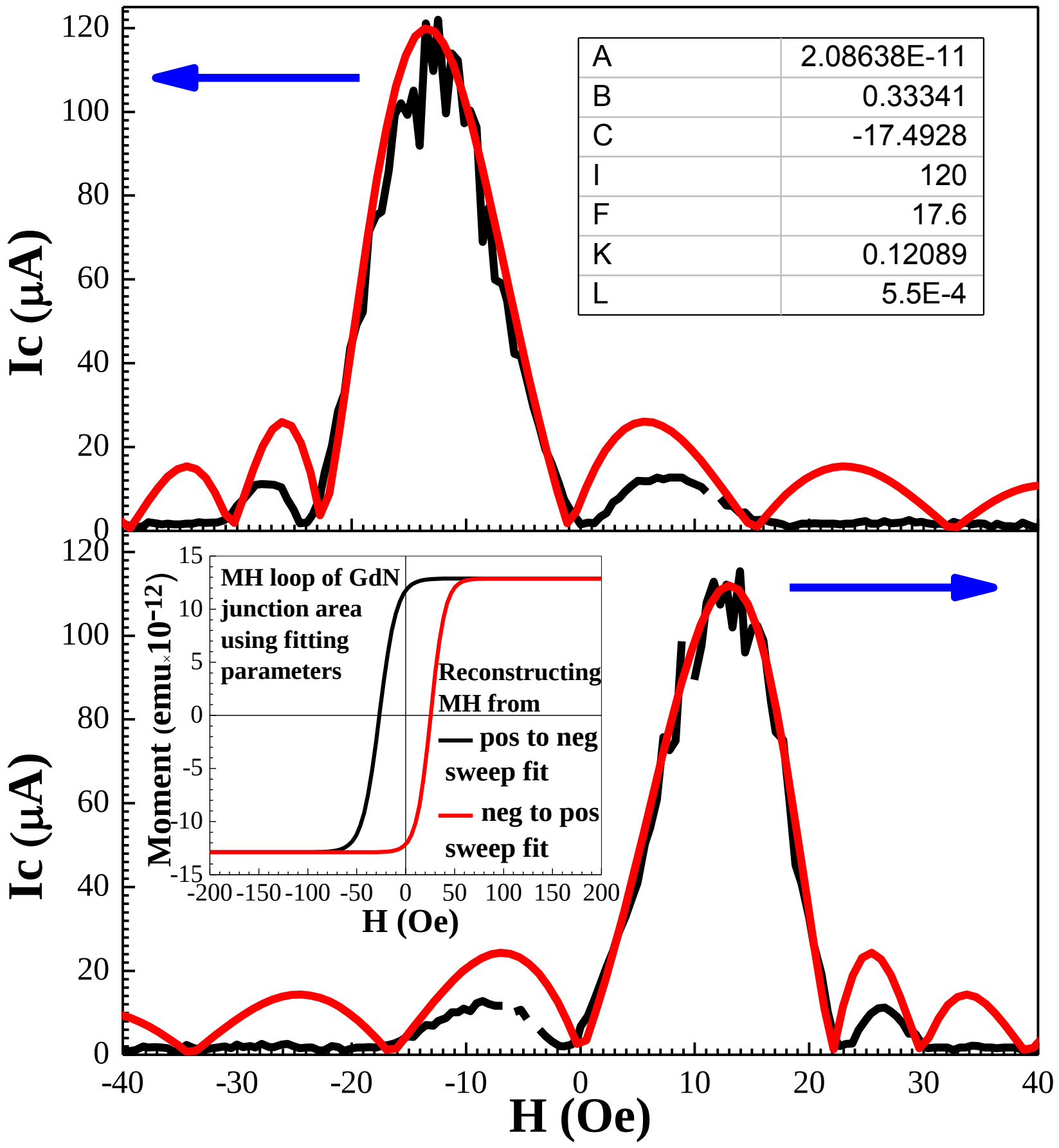

**FIG 2** – Fit to data of Figure 1(b) and 1(c) using the expressions derived in the text which take into account the role of barrier magnetism in $I_C(H)$ patterns. The matching fits in the horizontal axes agrees with our assumption that barrier magnetism distorts the patterns from their conventional Fraunhofer type shapes and that $H_{2+}$ is least distorted due to magnetic effects. Inset shows the reconstruction of M-H loops of the device area of GdN, the parameters for which are obtained from the values used for fitting the $I_C(H)$ patterns.

Fig 2 shows the fitting performed on data presented in Figure 1b, 1c by using the above mentioned equations. It is evident that the second lobe on the positive axes during the positive to negative field sweep (or second lobe on negative field axes during negative to positive field sweep) is least distorted. From the fits shown above, we find that when compared to the expected lobe size ($F$), $H_{2+}$ shows a distortion of only about 5%, whereas $H_1$ and, particularly, $H_{2-}$ are distorted considerably more. Thus, we use the experimentally obtained values of $H_{2+}$ as the most reliable measure of period of $I_c(H)$.

In Fig 3, we compare $I_c(H)$ patterns of magnetic and non-magnetic junctions. NbN/ MgO/ NbN junctions and NbN/ GdN/ NbN junctions with a non-magnetic GdN (low thickness of GdN barrier, equivalent to dead layer thickness) have almost identical characteristics expected of a $I_C = I_0 \left|\frac{sin(\pi\Phi/\Phi_0)}{(\pi\Phi/\Phi_0)}\right|$ dependence, where $\Phi_0$ is the flux quantum, $\Phi = HL(2\lambda + d)$, $L$ is the junction length perpendicular to applied field, $\lambda$ is the London penetration depth, $d$ is the GdN thickness. It is evident that the various parameters defined in Fig.1 in order to characterize the $I_c(H)$ patterns are strongly affected by increasing the GdN barrier thickness, which in turn increases the barrier magnetism and $P$.

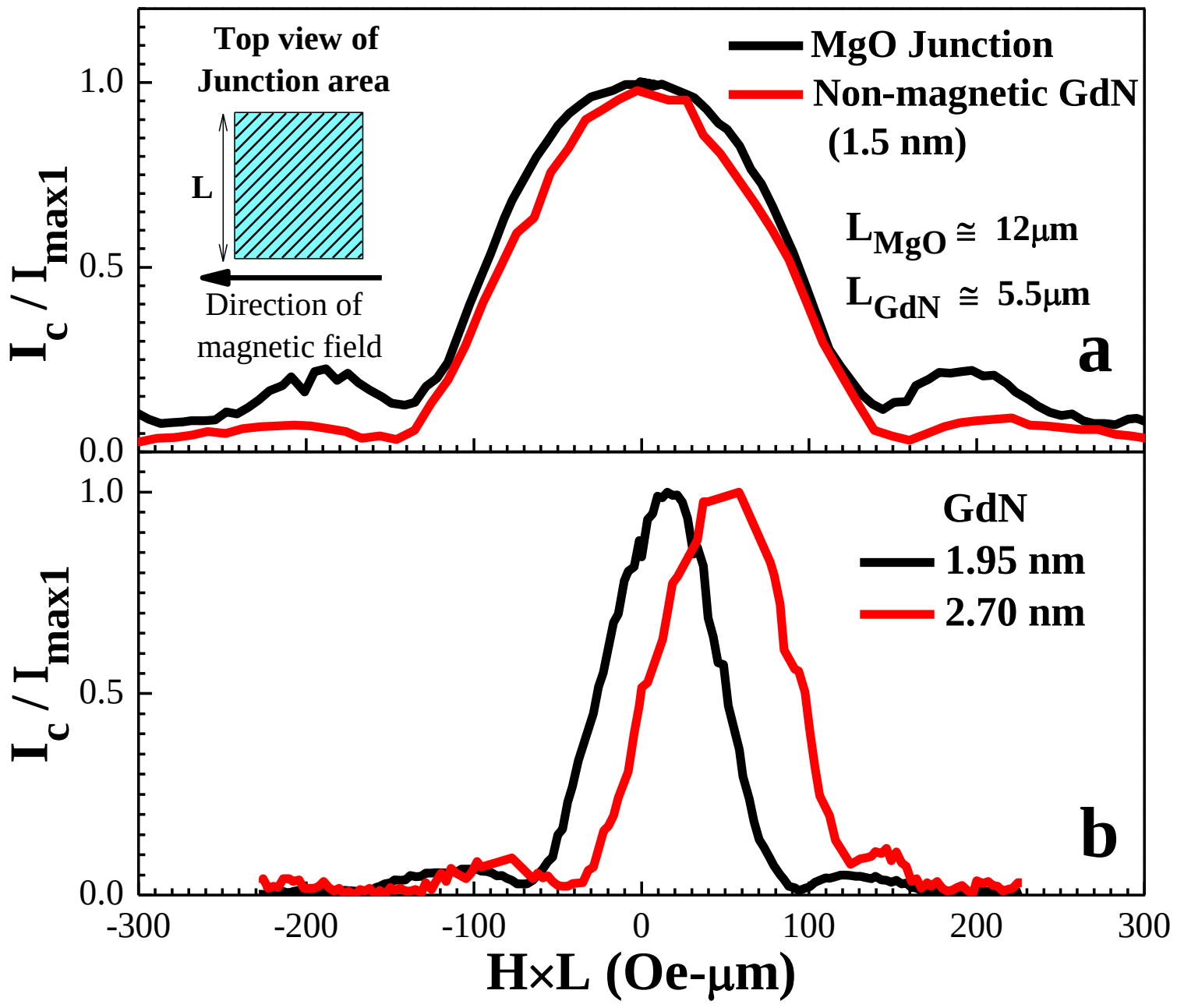

**FIG 3 –** (a) NbN/ MgO/ NbN junction of different geometry compared to a non-magnetic GdN junction, to establish that the low thickness non-magnetic GdN and a non-magnetic insulator are similar. (b) Magnetic interference patterns of JJs with 2

different thickness of GdN layer which demonstrate how parameters described in Fig.1 evolve with increasing GdN thickness. The observed differences in characteristics of $\boldsymbol{I_c(H)}$ patterns can thus be attributed to magnetism of GdN barrier.

In Fig 4 we plot various junction parameters as a function of GdN thickness. Most data points are averages of at least 4 junctions of the same barrier thickness and the error bars indicate the spread of values. The critical current is measurable up to a thickness of 2.9 nm but the second lobe in the $I_c(H)$ patterns is undetectable above 2.7 nm.

The primary results of this paper are reported in Fig 4a, b which show the changes in the $I_c(H)$ lobe-width with GdN thickness. The lines labelled $H_0$ in Fig 4a, b are derived from all the non-magnetic junctions in our study where $H_0 = \Phi_0/(L(2\lambda + d))$ – i.e. the lobe width of a conventional junction.

The NbN electrodes are of identical thickness for all samples, the device dimensions are constant and for NbN films $\lambda$ is >> d and hence $H_1$ and $H_{2+}$ should be equal to $H_0$ regardless of the GdN thickness. However, it is evident that for the junctions in which the magnetism and spin polarisation are well-developed, $H_1$ and $H_{2+}$ converge rapidly to a value of $H_0/2$ once magnetism is established. Because the other junction parameters are fixed, this observed decrease in $H_1$ and $H_{2+}$ therefore has a more fundamental origin. Earlier it was argued that $H_{2+}$ gives the most accurate estimate of the undistorted magnitude of the $I_c(H)$ lobes, and it clear from the data that the deviation from $H_0/2$ is less than 10%; i.e. for fully magnetic barriers the $I_C(H)$ patterns evolve to have exactly half the period expected from a conventional $I = I_C|sinx/x|$ relation; a behaviour expected of junctions which have CPR dominated by second harmonic[7,11]. An identical trend is found for behaviour of $H_1$ albeit with a somewhat larger scatter arising from the distorting effects arising from the reversal of GdN magnetism. Once established, this halving of the period is independent of GdN thickness; in contrast, $H_{2-}$ lies in the field region in which the barrier moment is switching and so, with increasing barrier thickness, $H_{2-}$ is progressively reduced as shown in Fig 4a.

This interpretation of second harmonic-dominated CPR for the most magnetic barriers is supported by the data in Fig 4 (c), where the ratio of $I_{max2}/I_{max1}$ initially decreases from the value of 0.21 expected for the standard Fraunhofer pattern as the magnetism turns on and the recovers again ~0.21 for the most magnetic barriers implying that, provided the hysteretic effect of the barrier flux is accounted for, these junctions show ideal half-period $I_c(H)$ patterns as predicted elsewhere[7]. We have attempted to measure Shapiro steps in these devices, but have been unable to couple microwave power into them.

The hysteresis ($\Delta H$) in $I_c(H)$ shown in Fig 4e increases linearly with thickness as expected if the barrier moment couples directly with the externally applied magnetic field. The spin polarization at 15K ($P_{15K}$) shown in Fig 4e shows a similar trend with thickness as observed previously[9].

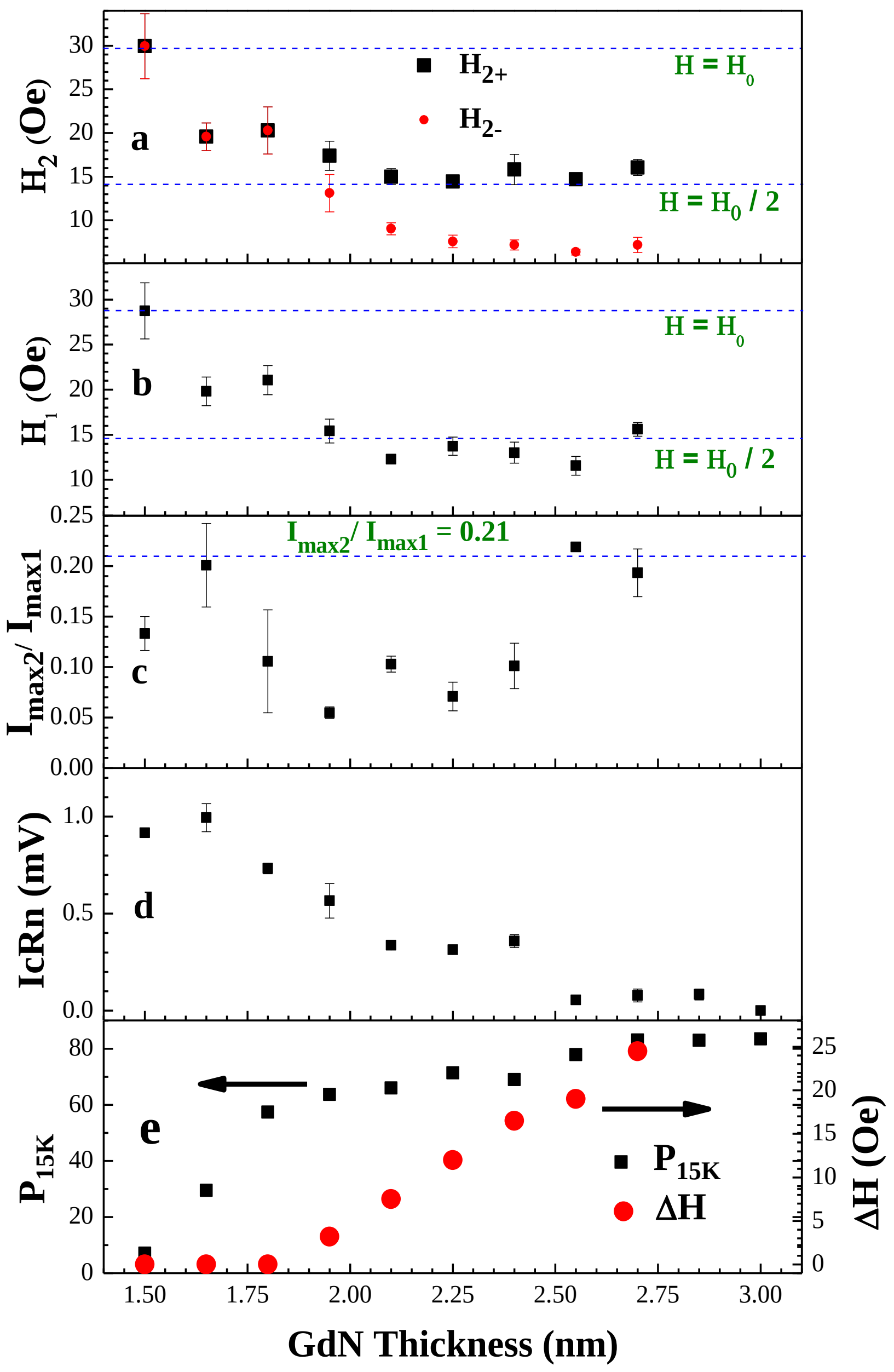

**FIG 4 –** Plot of thickness variation of various junction parameters. (a) Magnitude of second lobes $\boldsymbol{H_2}$ on either side (b) Magnitude of first lobe $\boldsymbol{H_1}$ (c) Ratio of $\boldsymbol{I_{max2}/I_{max1}}$ (d) $\boldsymbol{I_cR_n}$ product (e) Hysteresis in $\boldsymbol{I_c(H)}$ (f) Spin polarization at 15K .

## Discussion

Our results provide direct evidence for a pure 2$^{nd}$ harmonic in the current-phase-relation of NbN/GdN/NbN devices for all barrier thicknesses greater than 2 nm. Since, the second harmonic in our devices is independent of barrier thickness and in the sense originally introduced by Trifunovic[5], the dominance of the 2$^{nd}$ harmonic is *robust;* meaning its origin cannot be explained on the basis that the devices are at or near $0-\pi$ transitions, as such regions are limited to narrow and specific barrier thickness[12]. Therefore, the 2nd harmonic reported here must have a more fundamental origin. In recent theory papers[5,6] it was shown that under circumstances in which the supercurrent originates from the coherent transport of two triplet Cooper pairs, the current phase periodicity is necessarily doubled. On the basis of these theories, the two most important ingredients for the appearance of a robust 2nd harmonic are present in our data. Firstly, the data in Fig 4(e) shows that the 2nd harmonic appears at high spin polarisation which can provide the necessary filtering out of conventional spin singlet Cooper pairs which gives rise to a dominant 1st harmonic. Secondly, from the quasiparticle conductance spectrum[13] of our devices we have already observed an asymmetry in the exchange fields appearing in the two electrodes which might be implicated in a spin-mixing process at the $S/I_F$ interfaces[14]. In the absence of an adequate theory of 2nd harmonics in the tunnelling limit, a direct comparison with theory is not possible; however, our results nevertheless strongly suggest that tunnelling in a spin-filter JJ is mediated by an Andreev bound state that is more complex than current theoretical pair tunnelling models predict[15].

## Methods

The NbN/GdN/NbN films were grown on oxidized Si substrates pre-coated with a 10-nm-thick layer of MgO; the MgO layer protects the oxidized Si during fabrication by acting as an etch stop layer while etching NbN with $CF_4$, and serves no other purpose in device function. The thicknesses of the top and bottom NbN layers were 90 nm and 100 nm respectively. NbN and GdN films were deposited by reactive d.c. sputtering in an Ar/$N_2$ atmosphere from pure Nb and Gd metal targets in an ultra-high vacuum chamber without breaking the vacuum. Multiple substrates were rotated at differing speeds below a stationary Gd target using a computer-controlled stepper-motor in order to obtain samples with different GdN barrier thicknesses in the same deposition run. The MgO barrier was deposited by radio frequency sputtering from an MgO target in pure argon, followed by a post-deposition rf plasma oxidation stage. Mesa tunnel junctions were then fabricated using a four-stage lithography process. Fabricated junctions were measured by a four-point current bias technique using a dip probe in a liquid helium dewar. A solenoid was used to apply in-plane magnetic fields. The sample space was shielded from external stray fields by means of a $\mu$-metal shield.

## Author contributions

A.P prepared the GdN samples and carried out the experiments. Z.H.B prepared the MgO samples. A. P and M. G. B analysed the results. A. P, J.W.A.R and M.G.B prepared the manuscript.

## Acknowledgements

'This work was supported by the ERC Advanced Investigators Grant “SUPERSPIN” and EPSRC Grant EP/I036303/1. A. P was funded by a scholarship from the Cambridge Commonwealth Trust. J. W. A. R acknowledges funding from the Royal Society through a University Research Fellowship.

## Additional information

The authors declare no competing financial interests. Supplementary information accompanies this paper. Correspondence should be addressed to A. P.